# A Verified Decision Procedure for Orders in Isabelle/HOL

Lukas Stevens and Tobias Nipkow

Fakultät für Informatik, Technische Universität München, Munich, Germany

**Abstract** We present the first verified implementation of a decision procedure for the quantifier-free theory of partial and linear orders. We formalise the procedure in Isabelle/HOL and provide a specification that is made executable using Isabelle's code generator. The procedure is already part of the development version of Isabelle as a sub-procedure of the simplifier.

## 1 Introduction

Powerful proof automation facilities, e.g. `auto` in Isabelle, are crucial to make an interactive theorem prover practical. These tools fill in the logical steps that are trivial to humans and thus enable the users of interactive theorem provers to write formal proofs that resemble the less formal pen-and-paper proofs. Their efficacy in an interactive environment is judged by their completeness ("How many problems do they solve?") and their performance ("How fast do they solve the problems?"). Many of theses problems are undecidable in general; hence, incomplete heuristics, which are fast in practice, are used to tackle them. In decidable theories we can do better since they admit decision procedures, i.e. methods that always prove or disprove the goal at hand. Nevertheless, theorem provers sometimes still employ heuristics even for decidable theories. Isabelle in particular uses an unverified and incomplete ML procedure[1], which interprets a given set of (in)equalities as a graph, to decide partial and linear orders. As an example, the procedure fails to prove the goal

   **lemma assumes** ¬ x < y **and** x = y **and** ¬ x ≤ y **shows** False where ≤ is a partial order. Note that ¬ x < y is equivalent to x ≠ y ∨ x ≤ y for partial orders ≤. With that in mind, we investigate partial and linear order relations and develop decision procedures for them, which we prove to be sound and complete.

### 1.1 Related work

The decidability of the first-order theory of linear orders was posed as a problem in a article by Janiczak [5] that was posthumously published. In a review of the article from 1954, Kreisel [6] answers the question by reducing the theory of linear order to first-order monadic predicate calculus assuming no limit points in the order. Janiczak had died in 1951, though, and Kreisel's proof apparently went unnoticed in the literature. Subsequently, Ehrenfeucht [3] submitted an abstract that proposed a proof using model-theoretic methods; however, the result was never published. The problem was settled by Läuchli and Leonard [7], who proved decidability by showing both the set of valid and refutable sentences to be recursively enumerable.

---

[1] File path of the procedure in the Isabelle2021 distribution: `src/Provers/order.ML`

More recently, Negri et al. [8] performed a proof-theoretical analysis of order relations in terms of a contraction-free sequent calculus. Their analysis showed that the quantifier-free theory of partial orders has the subterm property. Due to the nature of the calculus, it follows that the proof search is terminating thus yielding a decision procedure. The result also extends to linear orders.

### 1.2 Contributions

In this paper, we develop decision procedures for the quantifier-free theory of partial and linear orders and provide an executable specification in Isabelle/HOL [10], i.e. we can generate code from the specification using the code generator of Isabelle. More specifically, our procedure determines whether the conjunction $\bigwedge_{i=1}^{n} L_i$ is contradictory where each literal $L_i$ is an (potentially negated) atom of the form $x = y$, $x \leq y$ or $x < y$. Note that it is possible to generalise this to arbitrary propositional formulas $\phi$ by taking their disjunctive normal form (DNF) and applying the procedure to each clause: the formula $\phi$ is a contradiction if and only if all clauses of the DNF lead to a contradiction. The restriction to a single clause is reasonable because we integrated the procedure as a sub-procedure of the simplifier and by extension of the classical reasoning tactics of Isabelle. As they eliminate disjunctions by performing case distinctions, an explicit conversion to DNF is not necessary.

Unlike Negri et al. [8], whose proof-theoretic procedure is too far removed from an actual implementation to give an accurate bound on the running time, we provide an implementation of the decision procedure and state its complexity. We also define a proof system that provides us with a framework to certify any contradiction that the procedure finds. Soundness and completeness of the executable specification is fully verified in Isabelle/HOL.

The paper is structured as follows: we start by giving a formal semantics for order (in)equalities in Section 2. To determine whether a set of (in)equalities is contradictory, we present abstract decision procedures for partial and linear orders in Section 3 and 4, respectively. The executable specification presented in Section 6 uses the proof terms introduced in Section 5 to certify contradictions. The final Section 7 gives an overview of how the exported code obtained from the executable specification is used to implement a tactic that can be applied to proof goals in Isabelle.

A copy of the formalisation is available online [12].

### 1.3 Notation

Isabelle/HOL conforms to everyday mathematical notation for the most part. For the benefit of the reader that is unfamiliar with Isabelle/HOL, we establish notation and in particular some essential datatypes together with their primitive operations that are specific to Isabelle/HOL. We write `t :: 'a` to specify that the term `t` has the type `'a` and `'a ⇒ 'b` for the type of a total function from `'a` to `'b`. The types for booleans, natural numbers, and in integers are `bool`, `nat`, and `int`, respectively. Sets with elements of type `'a` have the type `'a set`. Analogously, we use `'a list` to describe lists, which are constructed as the empty list `[]` or with the infix constructor `#`, and are appended with the infix operator `@`. The function `set` converts a list into a set. For optional values, Isabelle/HOL offers the type `option` where a term `opt :: 'a option` is either `None` or `Some a` with `a :: 'a`. Finally, we remark that **iff** is equivalent to = on type `bool` and ≡ is definitional equality of the meta-logic of Isabelle/HOL, which is called Isabelle/Pure.

## 2  A Semantics for Orders

Since we only deal with a single conjunction $\bigwedge_{i=1}^{n} L_i$ of order literals, it is convenient to represent it in clause form, i.e. just as a set of literals. A literal consists of a boolean polarity and an order atom. If the boolean is `True`, they are called positive; conversely, when the boolean is `False`, we call them negative. Altogether we define the type of order atoms and literals as follows:

  `type_synonym var = int`

  `datatype order_atom = var ≤ var | var < var | var = var`

  `type_synonym order_literal = bool × order_atom`

The **boldface** symbols **≤**, **<**, and **=** are ordinary constructors of the datatype chosen to resemble the (in)equalities they represent. Depending on context, we will abuse the notation for literals, e.g. we use `x ≤ y` to mean `(True, x ≤ y)` and `x ≮ y` to mean `(False, x < y)`. We arbitrarily chose to represent variables with `int` for straightforward code generation but any linearly ordered type would do.

Semantically, an order literal corresponds to a proposition that the (in)equality holds. To this end, we assign each variable `x` a value `v x :: 'a` and interpret the literals relative to a relation `r :: 'a rel` where `'a rel = ('a × 'a) set`. This allows us to apply the semantics in the context of any order relation by suitably constraining `r`. For example, we would demand that `r` is reflexive, transitive, and antisymmetric in the context of partial orders. We call a pair `(r, v)` a *model* of a literal `a` if `(r, v) ⊨_o a`, as defined below, holds. If `r` is constrained to a specific kind of order relation, say a partial order, we will speak of a *partial order model* `(r, v)`. The same notation is used for sets of literals `A` where `(r, v) ⊨_o A` is equivalent to $\forall a \in A.\ (r, v) \vDash_o a$.

```
fun ⊨ₒ :: 'a rel × (var ⇒ 'a) ⇒ order_literal ⇒ bool where
  (r, v) ⊨ₒ (p, x ≤ y) = (p ⟷ (v x, v y) ∈ r)
| (r, v) ⊨ₒ (p, x < y) = (p ⟷ (v x, v y) ∈ r ∧ v x ≠ v y)
| (r, v) ⊨ₒ (p, x = y) = (p ⟷ v x = v y)
```

## 3  Deciding Partial Orders

In this section, we will derive an abstract specification for a procedure that decides the theory of partial orders. For the time being, we assume that the set of (in)equalities `A` does not contain any strict inequalities and instead deal with them later in Section 5.2. In order to satisfy the (in)equalities, we need to come up with a partial order `r` and a variable assignment `v` that are a model of `A`, i.e. `(r, v) ⊨ₒ A`. We build a syntactic model where the variable assignment maps every variable to itself, i.e. we set `v = (λx. x)`. To find the accompanying relation, we first define a relation `leq1 A` that contains all pairs that are directly given by the set of (in)equalities; for example, we add `(x, y)` if `x ≤ y ∈ A`. A partial order has to be reflexive and transitive so we define `leq A` as the smallest reflexive and transitive relation that contains `leq1 A`.

```
definition leq1 A ≡ {(x, y). x ≤ y ∈ A ∨ x = y ∈ A ∨ y = x ∈ A}
definition leq A ≡ (leq1 A)*
```

Since we chose `v` to be the identity, we can directly read off all inequalities that must hold from `leq A`. Furthermore, we can use the same inequalities to derive all equalities that must hold by antisymmetry: the equality `x = y` must hold if both `(x, y) ∈ leq A` and `(y, x) ∈ leq A`. Bossert and Suzumura [2] call this subset of a relation the symmetric factor.

```
definition sym_factor r ≡ r ∩ r⁻¹
```

The symmetric factor is clearly a symmetric relation. Considering that reflexivity and transitivity is invariant under inversion and intersection, we conclude that `sym_factor r` is an equivalence relation for any relation `r` that is reflexive and transitive. The symmetric factor of `leq A`, called `eq A`, is thus an equivalence relation.

```
abbreviation eq A ≡ sym_factor (leq A)
```

Equipped with the relations `leq A` and `eq A`, we are ready to define the abstract decision procedure. It uses these relations, which are derived from the positive literals in `A`, and checks for consistency with the negative literals in `A`.

```
definition contr :: order_literal set ⇒ bool where
contr A ⟷ (∃ x y. x ≰ y ∈ A ∧ (x, y) ∈ leq A) ∨
           (∃ x y. x ≠ y ∈ A ∧ (x, y) ∈ eq A)
```

We claim that `contr` is a decision procedure for the theory of partial orders. To verify this claim, we have to prove soundness and completeness of `contr` with respect to our semantics ⊨ₒ. More precisely, we have to show that `contr A` evaluates to `True` if and only if `A` is contradictory, i.e. there exists no partial order model `(r, v)` of `A`. We prove both directions in contrapositive form, starting with soundness.

```
theorem contr_sound:
  assumes refl r and trans r and antisym r and (r, v) ⊨_o A
  shows ¬ contr A
```

The soundness proof, which is sketched below, uses the following lemma.

```
lemma assumes (x, y) ∈ leq A and (r, v) ⊨_o A shows (v x, v y) ∈ r*
```

*Proof (Soundness).* We assume `contr A` and therefore must show `False`. It holds by definition of `contr A` that a negative literal in `A` contradicts with either `leq A` or `eq A`.

We first consider the case where $x \not\leq y \in$ `A` and `(x, y) ∈ leq A` for some `x` and `y`. Using the assumption `(r, v) ⊨_o A`, we can apply the above lemma to conclude that `(v x, v y) ∈ r*`. Moreover, we have that `r* = r` because we assumed that `r` is reflexive and transitive. But this is a contradiction to the assumption `(r, v) ⊨_o A` which requires `(r, v) ⊨_o (x ≰ y) ⟷ (v x, v y) ∉ r` to hold.

In the remaining case we have $x \neq y \in$ `A` and `(x, y) ∈ eq A`. Remember that we defined `eq A` as the symmetric factor of `leq A` which implies that `(x, y) ∈ leq A` and `(y, x) ∈ leq A`. With the same argument as above it follows that `(v x, v y) ∈ r` and `(v y, v x) ∈ r`, and, by antisymmetry, `v x = v y`; however, this contradicts the assumption `(r, v) ⊨_o A` which entails `(r, v) ⊨_o (x ≠ y) ⟷ v x ≠ v y`. □

For completeness, on the other hand, we have to show that there exists a partial order model `(r, v)` for `A` if ¬ `contr A`. A tempting candidate for `r` would be `leq A` but then again `leq A` is only a preorder: it is not antisymmetric because it captures the relation between distinct variables, not their values. This means that `v` must map distinct variables `x` and `y` to the same value `v x = v y` if `(x, y) ∈ leq A` and `(y, x) ∈ leq A`. In other words, we have to take the quotient set of `leq A` with respect to the equivalence relation `eq A`.

Viewing `leq A` more abstractly as a preorder `r` on some set `C`, we have to map each variable `x` to its equivalence class in the equivalence relation `sym_factor r`. Lifting `r` to the quotient set `C // sym_factor r`, where `//` is Isabelle notation for the quotient, yields an antisymmetric relation and therefore a partial order.

```
definition sym_class r x ≡ {y | (x, y) ∈ sym_factor r}
definition sym_class_rel r ≡ {(sym_class r x, sym_class r y) |
                              (x, y) ∈ r}
```

We confirm that the lifting works as intended with the following lemma.

```
lemma assumes preorder_on C r
      shows (x, y) ∈ r ⟷
            (sym_class r x, sym_class r y) ∈ sym_class_rel r
```

Now, we apply these ideas to the relation `leq A` to obtain the partial order `Leq A`. Additionally, we take the reflexive closure of `Leq A` to obtain a partial order on the whole universe `UNIV` of the type `var set`.

**abbreviation** Eq :: (bool × atom) set ⇒ var ⇒ var set
  **where** Eq A x ≡ sym_class (leq A) x

**abbreviation** Leq :: (bool × atom) set ⇒ var rel
  **where** Leq A ≡ sym_class_rel (leq A)

**abbreviation** Leq_refl ≡ (Leq A)⁼

We show that, if ¬ `contr A`, then the interpretation (`Leq_refl A`, `Eq A`) is a model of A thus proving completeness of `contr`.

**theorem** contr_complete:
  **assumes** ¬ contr A **shows** (Leq_refl A, Eq A) ⊨$_o$ A

*Proof.* We show that for any `a ∈ A` it holds that (`Leq_refl A`, `Eq A`) ⊨$_o$ a. By case distinction on `a`, we prove that the statement holds for any kind of literal `a`. The proofs of the different cases are similar so we only present the case `a = x ⩽̸ y` for some `x` and `y`. Considering the definition of `contr` it follows from the assumption ¬ `contr A` that (x, y) ∉ `leq A`. This means that x ≠ y because `leq A` is reflexive. Moreover, we apply the above lemma to obtain (`Eq A x`, `Eq A y`) ∉ `Leq A`. Again, `Leq A` is reflexive on `UNIV // eq A` so we have `Eq A x` ≠ `Eq A y`. Since `Leq_refl A` only adds reflexive pairs to `Leq A`, we can conclude that (`Eq A x`, `Eq A y`) ∉ `Leq_refl A`. This gives us our goal (`Leq_refl A`, `Eq A`) ⊨$_o$ x ⩽ y ⟷ (`Eq A x`, `Eq A y`) ∉ `Leq_refl A`.  □

## 4 Deciding Linear Orders

We now show that the procedure `contr` can be modified to decide linear orders. Recall that the soundness of `contr` assumes that the underlying relation `r` is a partial order. This means that the soundness of `contr` for linear orders is trivial as every linear order is a partial order. Again, completeness is more involved since we have to construct a linear order model for A if ¬ `contr A` holds. Thus, we cannot reuse `Leq_refl A` because it is only a partial order. All is not lost, though: we can appeal to a classical result from order theory, namely Szpilrajn's extension theorem [13]. The original theorem states that every relation that is transitive and asymmetric, i.e. is a strict partial order, can be extended to a relation that is also total. A more general version of the theorem, which in particular applies to non-strict partial orders, was formalised in Isabelle/HOL by Zeller and Stevens in an AFP entry [14]. Using this result, we can prove that every partial order can be extended to a linear order.

**theorem** partial_order_extension:
  **assumes** partial_order r **shows** ∃ R. linear_order R ∧ r ⊆ R

We use Hilbert's ε-operator in the form of SOME to obtain an arbitrary extension Leq_ext A of the partial order Leq_refl A.

**definition** Leq_ext A ≡ (SOME r. linear_order r ∧ Leq_refl A ⊆ r)

Theorem partial_order_extension guarantees that such an extension exists so Leq_ext A is well-defined.

**theorem** linear_order (Leq_ext A) **and** Leq_refl A ⊆ Leq_ext A

Both Leq_refl A and Leq_ext A are reflexive and antisymmetric, which means that for any x, y with Eq A x = Eq A y we have

(Eq A x, Eq A y) ∈ Leq_refl A ⟷ (Eq A x, Eq A y) ∈ Leq_ext A.

Therefore, Leq_ext A is consistent with negative literals of the form x ≠ y ∈ A. The other case, that is to say literals of the form x ≰ y ∈ A, is not so easy: how can we ensure that the extension from Leq_refl A to Leq_ext A does not introduce (Eq A x, Eq A y) ∈ Leq_ext? Fortunately, we can sidestep the problem by exploiting the properties of linear orders: for any linear order r it holds that

$$
\begin{aligned}
(r, v) \vDash_o (x \not\leq y) &\longleftrightarrow (v\ x, v\ y) \notin r \\
&\longleftrightarrow v\ x \neq v\ y \land (v\ y, v\ x) \in r \\
&\longleftrightarrow (r, v) \vDash_o x \neq y, y \leq x.
\end{aligned}
$$

In other words, we can replace all literals of the form x ≰ y by the two literals x ≠ y and y ≤ x while maintaining the completeness of contr. Employing this preprocessing step, we obtain a sound and complete decision procedure for linear orders.

## 5 Certification with Proof Terms

We aim to generate an executable specification of the decision procedure in order to automatically generate code from it; however, we do not want to trust the code generation facilities of Isabelle. Instead, the executable specification has to certify any contradiction it finds with a proof term, which is then replayed through Isabelle's inference kernel to obtain a theorem.

### 5.1 Basic Proof System for Partial Orders

The proof system we define is very limited as the only kind of provable propositions are order literals. Since we chose to define order literals without an explicit constructor representing the boolean value False but ultimately want to deduce a contradiction, we define False in terms of the order literal 0 ≠ 0, that is Fls ≡ 0 ≠ 0. With this, we define the proof system ⊢$_P$ for partial orders (see Figure 1), where A ⊢$_P$ p : l means that the proof term p proves the proposition l under the set of assumptions A. For now, we will pretend that proof terms are defined as a datatype with one constructor for each rule as shown below. We will discuss the actual definition of proof terms in Section 5.

$$\frac{\text{x} \leq \text{y} \in \text{A}}{\text{A} \vdash_P \text{AssmP (x} \leq \text{y)} : \text{x} \leq \text{y}} \text{Assm} \qquad \frac{}{\text{A} \vdash_P \text{ReflP x} : \text{x} \leq \text{x}} \text{Refl}$$

$$\frac{\text{A} \vdash_P \text{p1} : \text{x} \leq \text{y} \qquad \text{A} \vdash_P \text{p2} : \text{y} \leq \text{z}}{\text{A} \vdash_P \text{TransP p1 p2} : \text{x} \leq \text{z}} \text{Trans}$$

$$\frac{\text{A} \vdash_P \text{p1} : \text{x} \leq \text{y} \qquad \text{A} \vdash_P \text{p2} : \text{y} \leq \text{x}}{\text{A} \vdash_P \text{AntisymP p1 p2} : \text{x} = \text{y}} \text{Antisym}$$

$$\frac{\text{x} = \text{y} \in \text{A}}{\text{A} \vdash_P \text{EQE1P (x} = \text{y)} : \text{x} \leq \text{y}} \text{EqE1} \qquad \frac{\text{x} = \text{y} \in \text{A}}{\text{A} \vdash_P \text{EQE2P (x} = \text{y)} : \text{y} \leq \text{x}} \text{EqE2}$$

$$\frac{(\text{False, a}) \in \text{A} \qquad \text{A} \vdash_P \text{p} : (\text{True, a})}{\text{A} \vdash_P \text{ContrP (False, a) p} : \text{Fls}} \text{Contr}$$

**Figure 1.** The proof system $\vdash_P$ for partial orders

```
datatype prf_trm = AssmP order_literal | ReflP order_literal |
  TransP prf_trm prf_trm | AntisymP prf_trm prf_trm |
  EQE1P order_literal | EQE2P order_literal |
  Contr order_literal prf_trm | ...
```

Every proof rule corresponds to a step the procedure `contr` takes:

- The relation `leq1 A` contains those pairs (x, y) for which we can prove `A ⊢_P p : x ≤ y` directly by assumption using one of Assm, EqE1, or EqE2.
- We obtain `leq A` from `leq1 A` by taking the reflexive transitive closure. Put another way, (x, y) ∈ `leq A` holds if and only if `A ⊢_P p : x ≤ y` can be proved proved by repeatedly applying the rules Refl and Trans.
- Since `eq A` is the symmetric factor of `leq A`, it contains exactly those pairs (x, y) for which `A ⊢_P p : x = y` is provable by the rule Antisym.
- Finally, we check if the negative literals are consistent with the relations `leq A` and `eq A`. Any inconsistency can be certified by the rule Contr.

Due to this close correspondence it is not surprising that we can prove the following lemmas.

```
lemma (x, y) ∈ leq A  ⟷  ∃p. A ⊢_P p : x ≤ y
lemma (x, y) ∈ eq A   ⟷  ∃p. A ⊢_P p : x = y
```

Using these lemmas, the soundness and completeness of the proof system relative to `contr` — and by extension to $\vDash_o$ — follow easily.

```
theorem ⊢_P_sound: assumes A ⊢_P p : Fls shows contr A
theorem ⊢_P_complete: assumes contr A shows ∃p. A ⊢_P p : Fls
```

## 5.2 Dealing with strict literals through rewriting

Until now, we assumed that the set of literals `A` does not contain any strict literals, i.e. literals of the form `x < y` or `x ≮ y`. For the case of linear orders `r`, dealing with those literals is just a matter of replacing them by equivalent, non-strict literals:

- $(r, v) \vDash_o (x < y) \longleftrightarrow (v\ x, v\ y) \in r \land v\ x \neq v\ y$
  $\longleftrightarrow (r, v) \vDash_o x \leq y,\ x \neq y$
- $(r, v) \vDash_o (x \not< y) \longleftrightarrow (v\ x, v\ y) \notin r \lor v\ x = v\ y$
  $\longleftrightarrow (v\ y, v\ x) \in r \lor v\ x = v\ y$
  $\longleftrightarrow (r, v) \vDash_o (y \leq x)$

Now, if `r` is a partial order, we can do the same for the former case but in the latter case we are stuck after the first step: $(v\ x, v\ y) \notin r \leftrightarrow (v\ y, v\ x) \in r$ does not hold. A possible solution is that we (recursively) check whether both `contr ({x ≰ y} ∪ A - {x ≮ y})` and `contr ({x = y} ∪ A - {x ≮ y})` hold. This was the first approach we tried but we ultimately found that the matching proof rule made reasoning about the proof terms tedious. Following Nipkow [9], we took a more general approach and introduced a type of propositional formulae with order literals as propositional atoms. Both replacement of literals and conversion to DNF are represented as rewrite rules on formulae. The semantics of order literals naturally generalises to formulae but, for brevity, we forgo discussing how we proved the soundness of the rewrite rules with respect to the semantics. A formula is either an atom or one of the logical connectives conjunction, disjunction, or negation:

```
datatype 'a fm = Atom 'a |
  And ('a fm) ('a fm) | Or ('a fm) ('a fm) | Neg ('a fm)
```

Motivated by the need to replace the order literals in the formula by other literals respectively formulae, we define $\text{amap}_{\text{fm}}$ which allows us to apply a replacement function to all atoms of a formula.

```
fun amap_fm :: ('a ⇒ 'b fm) ⇒ 'a fm ⇒ 'b fm where
  amap_fm f (Atom a) = f a
| amap_fm f (And ϕ₁ ϕ₂) = And (amap_fm f ϕ₁) (amap_fm f ϕ₂)
| amap_fm f (Or ϕ₁ ϕ₂) = Or (amap_fm f ϕ₁) (amap_fm f ϕ₂)
| amap_fm f (Neg ϕ) = Neg (amap_fm f ϕ)
```

We now define a function `deless` for partial orders that transforms a strict literal into a formula without strict literals.

```
fun deless :: order_literal ⇒ order_literal fm where
  deless (x < y) = And (Atom (x ≤ y)) (Atom (x ≠ y))
| deless (x ≮ y) = Or (Atom (x ≰ y)) (Atom (x = y))
| deless a = Atom a
```

We use the rules of the proof systems in Figure 2 to certify the rewrite steps that $\text{amap}_{\text{fm}}$ `deless` performs. The proof system for formulae $\equiv_{\text{fm}}$ is parametrised

$$\frac{}{\text{LessLe : x < y} \equiv_a \text{And (Atom (x} \leq \text{y)) (Atom (x} \neq \text{y))}} \text{LessLe}$$

$$\frac{}{\text{NlessLe : x} \not< \text{y} \equiv_a \text{Or (Atom (x} \not\leq \text{y)) (Atom (x = y))}} \text{NlessLe}$$

$$\frac{\text{p : a} \equiv_a \phi}{\text{AtomConv p : Atom a} \equiv_{\text{fm}} \phi} \text{AtomConv} \qquad \frac{}{\text{AllConv :} \phi \equiv_{\text{fm}} \phi} \text{AllConv}$$

$$\frac{\text{p :} \phi \equiv_{\text{fm}} \psi}{\text{NegConv p : Neg } \phi \equiv_{\text{fm}} \text{Neg } \psi} \text{NegConv}$$

$$\frac{\text{bop} \in \{\text{And, Or}\} \quad \text{p1 :} \phi_1 \equiv_{\text{fm}} \psi_1 \quad \text{p2 :} \phi_2 \equiv_{\text{fm}} \psi_2}{\text{BinopConv p1 p2 : bop } \phi_1 \phi_2 \equiv_{\text{fm}} \text{bop } \psi_1 \psi_2} \text{BinopConv}$$

**Figure 2.** Proof system $\equiv_{\text{fm}}$ for conversions of formulae

by a proof system for atoms $\equiv_a$. Again, you may imagine that the datatype of proof terms has a constructor for each rule of the proof systems.

We use the above rules to define functions that produce a proof term for $\text{amap}_{\text{fm}}$ deless.

```
fun amap_fm_prf :: ('a ⇒ prf_trm) ⇒ 'a fm ⇒ prf_trm where
  amap_fm_prf ap (Atom a) = AtomConv (ap a)
| amap_fm_prf ap (And φ₁ φ₂) =
    BinopConv (amap_fm_prf ap φ₁) (amap_fm_prf ap φ₂)
| amap_fm_prf ap (Or φ₁ φ₂) =
    BinopConv (amap_fm_prf ap φ₁) (amap_fm_prf ap φ₂)
| amap_fm_prf ap (Neg φ) = ArgConv (amap_fm_prf ap φ)

fun deless_prf :: order_literal ⇒ prf_trm where
  deless (x < y) = LessLe
| deless (x ≮ y) = NlessLe
| deless_prf _ = AllConv
```

We can show that $\phi \equiv_{\text{fm}} \text{amap}_{\text{fm}}$ deless $\phi$ : $\text{amap}_{\text{fm}}\_\text{prf}$ deless_prf $\phi$ by a simple inductive proof. After this conversion, the resulting formula may contain disjunctions but our decision procedure can only deal with conjunctions; thus, we first have to compute the DNF of $\psi$ and apply the procedure to each clause. Certifying the conversion to DNF follows a similar approach to $\text{amap}_{\text{fm}}$ so we refer to the formalisation for the details. This conversion also eliminates negations in the formula by pushing them into the atoms. Now assume that we are given a formula $\phi$ in DNF without negations, we still need to apply the decision procedure to each clause of the formula. As conversions alone are not sufficient, we build a proof system for propositional logic on top of conversions. Similarly to the system $\equiv_{\text{fm}}$, the proof system $\vdash$ in Figure 3 is parametrised by a proof system for atoms $\vdash_a$ (with the same type as $\vdash_P$).

$$\frac{A \vdash_a \mathtt{p} : \phi}{A \vdash \mathtt{p} : \phi} \text{ Lift} \qquad \frac{\mathtt{And\ c\ d} \in A \quad A,\mathtt{c},\mathtt{d} \vdash \mathtt{p} : \phi}{A \vdash \mathtt{ConjE\ c\ d\ p} : \phi} \text{ ConjE}$$

$$\frac{\mathtt{Or\ c\ d} \in A \quad A,\mathtt{c} \vdash \mathtt{p1} : \phi \quad A,\mathtt{d} \vdash \mathtt{p2} : \phi}{A \vdash \mathtt{DisjE\ c\ d\ p1\ p2} : \phi} \text{ DisjE}$$

$$\frac{\mathtt{c} \in A \quad \mathtt{c} \equiv_{\mathrm{fm}} \mathtt{d} : \mathtt{cp} \quad A,\mathtt{d} \vdash \mathtt{p} : \phi}{A \vdash \mathtt{Conv\ cp\ p} : \phi} \text{ Conv}$$

**Figure 3.** Propositional proof system for formulae.

A clause $C$ of a formula in DNF consists of nested applications of the constructor `And` with `Atom` constructors as leaves. We first define a function `conj_list` that computes the atoms of $C$. Along with it, we define a function `from_conj_prf` that uses the rule ConjE to convert the proof `p` that assumes every atom in `from_conj` $C$ into a proof that just assumes $C$.

```
fun conj_list :: 'a fm ⇒ 'a list where
  conj_list (And φ₁ φ₂) = conj_list φ₁ @ conj_list φ₂
| conj_list (Atom a) = [a]

fun from_conj_prf :: prf_trm ⇒ 'a fm ⇒ prf_trm where
  from_conj_prf p (And a b) =
    ConjE a b (from_conj_prf (from_conj_prf p b) a)
| from_conj_prf p (Atom a) = p
```

Let `contr_prf`$_a$ `:: 'a list ⇒ prf_trm option` be a function that tries to derive a contradiction from a list of atoms. We will define an instance `contr_list` of `contr_prf`$_a$ that refines the abstract procedure `contr` in the upcoming section. In order to prove that $\phi$ is contradictory, we first recurse down to its clauses and apply `contr_prf`$_a$ to each clause. Then, if each clause is contradictory, we combine those inductively with the rule DisjE to obtain a proof for the whole formula.

```
fun contr_fm_prf :: 'a fm ⇒ prf_trm option where
  contr_fm_prf (Or c d) = case (contr_fm_prf c, contr_fm_prf d) of
    (Some p1, Some p2) ⇒ Some (DisjE c d p1 p2) | _ ⇒ None
| contr_fm_prf (And a b) = case contr_prfₐ (conj_list (And a b)) of
    Some p ⇒ Some (from_conj_prf p (And a b)) | None ⇒ None
| contr_fm_prf (Atom a) = contr_prfₐ [a]
```

To summarise, we now have the tools to preprocess a conjunction of partial order literals such that we can apply the decision procedure `contr` to the clauses of the resulting formula. By introducing appropriate proof terms, the same tools can be applied to linear orders where it is not necessary to convert to DNF. The functions as defined above are amenable to code generation; thus, the only missing part is an executable specification for `contr`, which is the topic of the next section.

# 6 Refinement to Executable Specification

The executable specification utilises the abstract datatype (`'a, 'b) mapping`, which is a partial map `'a ⇒ 'b option` from keys to values. Isabelle's library conveniently provides a refinement of `mapping` to red-black trees, thereby making `mapping` executable [4]. We will need the following operations on this datatype:

- `Mapping.keys m` give us all keys of the map `m` that have an associated value.
- `Mapping.entries m` gives us the entries of the map `m`, i.e. all key-value pairs.
- `Mapping.of_alist as` converts the association list `as` into a map.

The prefix `Mapping` will be dropped in what follows.

Remember that, at its core, the decision procedure computes the relation `leq A` where for each `(x, y) ∈ leq A`, there exists a proof `p` such that $A \vdash_P p : x \leq y$. Computing `leq A` boils down to computing the transitive closure of the finite relation `leq1 A` while keeping track of the corresponding proof terms. Note that we only assume finiteness for the sake of executability; the abstract decision procedure does not make this assumption. This in turn allows us to only consider a finite number of terms of `(leq1 A)`$^+$ = $\bigcup_{i=0}^{\infty}$ `(leq1 A)`$^{i+1}$. More specifically, if `leq1 A` contains `n` pairs, then it is sufficient to only consider the first `n` terms. We implement this naively by iterating over `n` while accumulating the `n`-fold relational composition. We claim (without formal proof) that this yields a running time of $\mathcal{O}($`n`$^4$ `log(n)`$)$ where the logarithmic component is due to the implementation being based on red-black trees. Using the Floyd-Warshall-Algorithm and arrays instead of red-black trees, the running time could be improved to $\mathcal{O}($`v`$^3) \subseteq \mathcal{O}($`n`$^3)$ where `v` is the number of distinct variables in the set of (in)equalities `A`. This optimisation, however, is unlikely to pay off since the goals tend to be small: throughout the whole basic library of Isabelle/HOL the number of order literals never exceeds 13. Although computing the transitive closure dominates the running time of the procedure, it must be noted that case analyses on literals of the form `x < y` incur an exponential number of calls to the procedure. Altogether we obtain a function `trancl_mapping` that computes the transitive closure

   **lemma assumes** finite (keys m)
        **shows** keys (trancl_mapping m) = trancl (keys m)

and keeps track of the proof terms:

   **lemma assumes** finite (keys m)
          **and** $\forall((x, y), p) \in$ entries m. $A \vdash_P p : x \leq y$
     **shows** $\forall((x, y), p) \in$ entries (trancl_mapping m). $A \vdash_P p : x \leq y$

As explained above we assume the set of order literals to be finite so we represent it as list. This makes defining an executable refinement for `leq1` straightforward. Here, computing the intermediate `leq1_list` is done strictly to simplify the proofs as one could use a fold over the mapping to obtain `leq1_mapping` directly.

```
fun leq1_member_list :: order_literal
                    ⇒ ((var × var) × prf_trm) list where
  leq1_member_list (x ≤ y) = [ ((x, y), AssmP (x ≤ y)) ]
| leq1_member_list (x = y) =
    [ ((x, y), EQE1P (x = y)), ((y, x), EQE2P (x = y)) ]
| leq1_member_list _ = []

definition leq1_list A ≡ concat (map leq1_member_list A)
definition leq1_mapping A ≡ of_alist (leq1_list A)
```

Equipped with the above functions, we can compute the transitive closure `trancl_mapping (leq1_mapping A)`; thus, we are only missing the reflexive closure to have a refinement of `leq`. The reflexive closure for an infinite carrier type, however, would yield an infinite set. Therefore, we only represent the set implicitly with the predicate `is_in_leq` that, for a given pair `(x, y)`, returns some proof $A \vdash_P p : x \leq y$ if and only if `(x, y) ∈ leq A`. Similarly, we define `is_in_eq` by combining the proofs we get from `is_in_leq` with the rule ANTISYM. We pass around `trancl_mapping (leq1_mapping A)` as the argument `leqm` to avoid recomputing it.

```
definition is_in_leq leqm (x, y) ≡
  if x = y then Some (ReflP x) else lookup leqm l

definition is_in_eq leqm (x, y) ≡
  case (is_in_leq leqm (x, y), is_in_leq leqm (y, x)) of
    (Some p1, Some p2) ⇒ Some (AntisymP p1 p2) | _ ⇒ None
```

Putting things together, we try to to find the first negative literal in `A` that stands in contradiction to either `is_in_leq` or `is_in_eq`. In case we find one, we produce a proof of contradiction by mapping `Contr` over the value with `map_option :: ('a ⇒ 'b) ⇒ 'a option ⇒ 'b option`.

```
fun contr1_list :: ((var × var), prf_trm) mapping ⇒ order_literal
                ⇒ prf_trm option where
  contr1_list leqm (x ≤ y) =
    map_option (ContrP (x ≤ y)) (is_in_leq leqm (x, y))
| contr1_list leqm (x ≠ y) =
    map_option (ContrP (x ≠ y)) (is_in_eq leqm (x, y))
| contr1_list _ _ = None

fun contr_list_aux where
  contr_list_aux leqm [] = None
| contr_list_aux leqm (l#ls) = case contr1_list leqm l of
    Some p ⇒ Some p | None ⇒ contr_list_aux leqm ls

definition contr_list A ≡
  contr_list_aux (trancl_mapping (leq1_mapping A)) A
```

The executable specification `contr_list` refines `contr`.

**theorem** contr (set A) ⟷ (∃p. contr_list A = Some p)

# 7 From Exported Code to Integrated Proof Tactic

In the previous sections, we demonstrated how to refine the abstract decision procedure down to an executable specification. We can now generate Standard ML code from it that, assuming the code generator to be correct, implements the specification. Nevertheless, the procedure only works on a simple term language of propositional formulas with order literals as their atoms. To integrate the procedure back into Isabelle as a full-blown tactic, we have to convert a goal given in the higher-order term language of Isabelle into our simple term language on the one hand and replay the proof terms produced by the procedure through Isabelle's inference kernel on the other hand. The first aspect is taken care of by some hand-written ML code that

- brings the goal into a form where we have to prove `False`,
- extracts those assumptions that are order literals,
- converts the literals into our simple representation, e.g. for the term `s ≤ t` it replaces ≤ by the constructor `≤` and `s` and `t` by integer variables,
- and builds conjunction from the converted literals using `And`.

Passing the conjunction of literals to the exported code produces a proof term in the format that we sketched in Section 5. There, we pretended that each rule of the proof system has a designated proof term constructor, which would require replay code for every constructor. In reality we use a more general format for the proof terms, namely a simplified version of Isabelle's proof terms as introduced by Berghofer and Nipkow [1]. Both terms and proof terms are less expressive in our representation. First, terms only consist of constants, function application, and variables, where each variable stands for an Isabelle term as explained above. In particular, there is no function abstraction.

```
datatype trm = Const String.literal | App trm trm | Var var
```

As for the proof terms, they are less expressive as well: we have proof constants, proof variables bound by an enclosing proof abstraction, proof application, proof abstraction, term application, and conversions but no term abstractions and no instantiation of proof constants. In contrast to Isabelle's proof terms, we refer to bound proofs by their proposition, i.e. with a term, instead of using variables. This is to avoid dealing with bound variable indices, which simplifies reasoning about proof terms. Conversion proofs are strictly for convenience as the other constructors would be sufficient to represent equational proofs.

```
datatype prf_trm = PThm String.literal | Bound trm
  | AppP prf_trm prf_trm | AbsP trm prf_trm
  | Appt prf_trm trm | Conv trm prf_trm prf_trm
```

In Figure 4, we define a proof system where we write $\Gamma \vdash \mathtt{p} : \phi$ to mean that, in the context $\Gamma$, the proof term `p` proves the proposition $\phi$. The context $\Gamma$ contains propositions and conversions but no terms because we omitted term abstractions. Quoting bound proof variables in $\Gamma$ by their term requires us to convert the simple terms to Isabelle terms. For this, we use the function `dr` that

$$\frac{\Sigma(\texttt{c}) = \phi}{\Gamma \vdash \texttt{PThm c} : \phi} \text{ PThm} \qquad \frac{}{\Gamma, \texttt{dr}(\phi) \vdash \texttt{Bound } \phi : \texttt{dr}(\phi)} \text{ Bound}$$

$$\frac{\Gamma, \phi \vdash \texttt{p} : \psi}{\Gamma \vdash \texttt{AbsP } \phi \texttt{ p} : \phi \Longrightarrow \psi} \text{ AbsP} \qquad \frac{\Gamma \vdash \texttt{p} : \phi \Longrightarrow \psi \quad \Gamma \vdash \texttt{q} : \phi}{\Gamma \vdash \texttt{AppP p q} : \psi} \text{ AppP}$$

$$\frac{\Gamma \vdash \texttt{p} : \bigwedge\texttt{x}.\ \phi}{\Gamma \vdash \texttt{Appt p t} : \phi[\texttt{dr(t)}/\texttt{x}]} \text{ Appt}$$

$$\frac{\texttt{rpc(cp)} = \texttt{cv} \quad \texttt{cv}(\texttt{dr}(\pi)) = (\texttt{dr}(\pi) \equiv \sigma) \quad \Gamma, \texttt{dr}(\pi), \sigma \vdash \texttt{p} : \phi}{\Gamma, \texttt{dr}(\pi) \vdash \texttt{Conv } \pi \texttt{ cp p} : \phi} \text{ Conv}$$

**Figure 4.** Proof system for proof terms

maps constants to Isabelle constants and variables back to their corresponding Isabelle terms. Proof constants `PThm c` are interpreted by the environment $\Sigma(\texttt{c})$, which maps them to the corresponding propositions. Finally, we use the function `rpc(cp)` to convert a conversion proof into an Isabelle conversion.

The rules for proof abstraction AbsP and for proof application AppP correspond to introduction respectively elimination of Isabelle's meta-implication $\Longrightarrow$. Similarly, applying a term to a proof with Appt is equivalent to elimination of the universal meta-quantification $\bigwedge$. Those rules are modelled after primitives of the Isabelle's inference kernel so they are straightforward to replay. The remaining rules, on the other hand, require retrieving information from the context $\Gamma$, which is implemented as follows: mapping from theorem and conversion constants `PThm c` to the respective theorems and conversions is realised with association lists. Likewise, we save the bound proof terms in a map from terms to theorems, recursively extending the map with assumptions introduced by AbsP while replaying the proof term. The function `rpc` that constructs a conversion from proof applications and conversion constants is straightforward to implement. Applying the resulting conversion to the specified bound proof term and adding the new theorem to the context is all we need to implement the rule Conv.

The procedure is already part of the development version of Isabelle[2] where it is registered to the simplifier as a so-called solver. As such, the procedure is called whenever the simplifier is out of applicable rewrite rules. This is helpful when, for example, the simplifier wants to apply a conditional rewrite rule whose premises talk about set inclusion (which is a partial order). Since our procedure is more powerful than the old one, more rewrite rules apply which resulted in some broken proofs that had to be fixed. There were no significant changes in performance in comparison to the old procedure.

## 8 Conclusion

We provided the first verified implementation of a decision procedure for the quantifier-free theory of partial and linear orders. Although we closely followed the Isabelle/HOL formalisation in our presentation, the findings are not specific to Isabelle: any reasonably powerful theorem prover could use the code exported from the specification and replay the proof terms that it produces. In future work, we plan to apply the methodology presented here to the quantifier-free theory of (reflexive) transitive closure. Another direction is to replace our terms and proof terms by those from the formalisation of Isabelle's meta-logic [11], allowing us to reason about higher-order terms directly instead of translating between them and our simple terms.

---

[2] Introduced in commit `https://isabelle-dev.sketis.net/rISABELLEa3cc9fa129`